# FTIR spectral imaging as a probe of ultrasound effect on cells *in vitro*.

L. Di Giambattista[1,2], P. Grimaldi[1], I. Udroiu[3], D. Pozzi[4], G. Cinque[5], M.D. Frogley[5], A. Giansanti[1] and A. Congiu Castellano[1].

[1] Physics Department, Sapienza, University of Rome (IT).
[2] CISB (Interdepartmental Research Centre for Models and Information Analysis in Biomedical Systems), Sapienza, University of Rome (IT).
[3] DIPIA, ISPESL, via Urbana 167, Rome (IT).
[4] Experimental Medicine and Pathology Department, Sapienza, University of Rome (IT).
[5] Diamond Light Source Ltd, Didcot, Oxfordshire (UK).

## Abstract

Safe and efficient intracellular delivery of genes or drugs is critically important in targeted cancer treatment and gene therapy applications. *Ultrasound* (US) has been demonstrated to alter the cell membrane permeability due to a biophysical mechanism (*Sonoporation*) and exploited as a promising non-invasive gene transfer method. The sonoporation process could induce the formation of transient pores without significantly affecting cell viability.

This research is aimed at investigating some bioeffects due to *Therapeutic Ultrasound* (pulsed-1 MHz) which could allow to enhance drugs or genes delivery in a non tumoral cell line. We have used the NIH-3T3 cell line as model system and exposed it to US at two different distances from the source; the effects of this pulsed ultrasonic wave on cells were assessed by *Fourier transform infrared* (FT-IR) spectroscopic imaging analysis. This technique combined with a *focal plane array* (FPA) detector has been widely used to study the general biochemical changes *in vitro*; moreover, the development of FPA detectors and shortening of measurement times specifically for IR imaging, from several hours to few minutes, have made possible to image the distribution of molecular species in biological samples. We have also performed a *cytokinesis-block micronucleus* (CBMN) assay to reveal the presence or not of *micronuclei* (named *Howell-Jolly bodies*) formed during the cell division due to the DNA damage.

The results of IR analysis combined with the cytogenetic analysis have shown that these experimental conditions can not cause DNA mutations in the NIH-3T3 cell line. Finally, the comparison between the spectral parameters of the average spectrum extracted from the spectral map and those of the set of all spectra from the spectral map could be limited by the presence of bad pixels inside the map.

## *List of symbols*

| Symbol | Description | Symbol | Description |
|---|---|---|---|
| $A_1$ | Amide I (-C=O stretching), proteins | $N_{bs}$ | Number of bad spectra |
| $A_2$ | Amide II (-N-H bending, -C-N stretching), proteins | MCT | Mercury cadmium telluride |
| BNC | Binucleated cells | MN | Micronucleus |
| CBMN | Cytokinesis-block micronucleus | MNC | Mononucleated cells |
| CTR | Control cells (untreated cells) | PBS | Dulbecco's Phosphate Buffer Saline |
| DAPI | 4'-6'-diamidino-2-phenylindole | PCA | Principal Component Analysis |
| DMEM | Dulbecco's Modified Eagle's minimum essential Medium | PCs | Principal components |
| FBS | Fetal serum bovine | PC1 | First principal component |
| FPA | Focal plane array | PC2 | Second principal component |
| FT-IR | Fourier transform infrared | $R_I$ | $\frac{A_{1I}}{A_{2I}}$ intensity ratio |
| HCA | Hierarchical Cluster Analysis | | |
| H | Inter-spectral distance | RMS | Root Mean Square |
| h | Intra-cluster heterogeneity | $SON_{t\_d}$ | Sonicated cells (treated cells), t=time and d=distance |
| $h_D$ | Distance between petri dish and transducer | S/N | Signal-to-Noise ratio |
| NDI | Nuclear division index | US | Ultrasound |
| NIH-3T3 | Mouse fibroblast cell line | | |

# 1. Introduction

In the last years, Ultrasounds have been employed in gene delivery for the advantages over other systems such as virus or nonvirus-mediated systems. Among its advantages there is the fact that it is a less invasive-method, with an a higher delivery efficiency, and more applications and minimal cell death can be achieved. US systems are available for both *in vitro* and *in vivo* studies [1-4].

US effects (*thermal, cavitation and microstreaming*) depends on physical and biological factors, such as frequency, intensity, time exposure, duty cycle, temporal and spatial structure of sound field, the physiological state and the size-volume of a sonicated sample, and external conditions like temperature, pressure. Such a great number of variables complicates the analysis of the phenomena [5]. The focus of this contribution is on the quantitative evaluation of collateral biological effects of the treatment of non tumoral cell lines.

We have used FT-IR spectroscopy imaging based on the FPA detector [6,7] to study US-bioeffects on a cellular system in the range of therapeutic ultrasound (1 MHz in pulsed system). This technique has been combined with a cytogenetic analysis (CBMN), that allows to study DNA-damage at chromosomic level. A micronucleus is formed during the metaphase/anaphase transition of mitosis; it may arise from a whole lagging chromosome (aneugenic event leading to chromosome loss) or an acentric chromosome fragment detaching from a chromosome after breakage (clastogenic event) which do not integrate in the daughter nuclei. These micronuclei are also known as *Howell-Jolly bodies*. *In vitro*, the analysis of cells in presence of *cytochalasin B* (added 44 hours after the start of cultivation), an inhibitor of actins, allows to distinguish easily between mononucleated cells which did not divide and binucleated cells which completed nuclear division during *in vitro culture*; the frequencies of mononucleated cells provide an indication of the background level of chromosome/genome mutations accumulated *in vivo*. Moreover, the frequencies of binucleated cells with MN allow to measure the damage accumulated before cultivation plus mutations expressed during the first *in vitro* mitosis.

According to the previous test, we have established the following conditions of experimental set-up: the size of the plexiglass tank, the distances and position between the cellular sample and transducer (*ultrasound source*) within tank, the use of US in pulsed system with 75 % of duty cycle.

Finally, we have compared the utility of different methods to evaluate the quality of FT-IR maps and the chemical similarity between the samples.

# 2. Materials and Methods

*2.1 Cell culture*

In this research, we have used a healthy adherent fibroblast cell line, named NIH-3T3; the NIH-3T3 were grown with a solution of *Dulbecco's Modified Eagle's minimum essential Medium* (DMEM) without calcium with 10% *fetal bovine serum* (FBS), 1% *penicillin-streptomycin* and 1% *L-glutamine* at 37° C in humidified atmosphere containing 95% air and 5% $CO_2$.

*2.2 Sample preparation*

Before the US exposure, the NIH-3T3 were cultured as monolayer on $CaF_2$ windows, pre-treated with polylysine, inside a 35-mm petri dish. The viability of both the untreated (*control*) and the US-treated (*sonicated*) cells has been determined by Trypan blue exclusion test (over 90% of viability). After the treatment with US, the control and sonicated cells were fixed in *paraformaldehyde* (2% for 15 min), washed in *Dulbecco's Phosphate Buffer Saline* (PBS) and in distilled water to remove PBS residues, and dried in a desiccator.

*2.3 Ultrasound exposure set up*

We have built a plexiglass tank filled with partially degassed water and the US transducer was placed at the bottom. According to preliminary tests, the following parameters were chosen: distances ($h_D$) between the cellular sample in a petri dish and the transducer were 10 and 15 cm (Figure 1), the times of US exposure were 15, 30, 45, 60 minutes.
Acoustic pressure at both distances was measured by a needle hydrophone (Precision Acoustics LTD, HP 0.5 mm Interchangeable Probe) with a sensitivity of -272.7dB re 1V/μPa at the frequency of 1 MHz (Table *a*) [8].

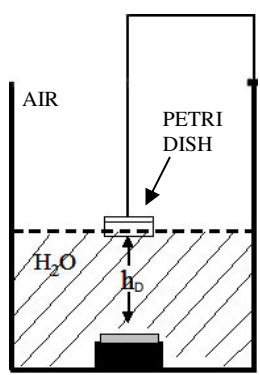

**Figure1.** *The general scheme of the experimental set-up.*

The US frequency has been set to 1 MHz with 100% of maximum power in pulsed system with 75% duty cycle.

| | 100% Power with 75% of duty cycle | |
|---|---|---|
| Distance (cm) | Pressure ± ΔP (Pa) | Acoustic Intensity ± ΔI (W/cm$^2$) |
| 10 | 214.5 ± 21 | $(3.11 ± 0.4)*10^{-6}$ |
| 15 | 318.7 ± 32 | $(6.86 ± 0.9)*10^{-6}$ |

**Table a.** *Values of the pressure within the petri dish and acoustic instant intensity at 10 and 15 cm.*

The temperature inside and outside of the petri dish at two different distances was measured; a maximum increase of 2 °C or less, that can not induce any breaks in the integrity of plasma membrane, was revealed during time of US exposure.
In the results, we have indicated the control cells as "CTR" and the sonicated cells as "SON**t_d**" where **t**=0,15,30,45,60 minutes and **d**=10,15 cm.

*2.4 Measurement using FPA detector*

An FTIR microscope (Hyperion 3000) coupled to an Vertex 80v FTIR spectrometer (both from Bruker Optics, Germany) was used to the analysis and was controlled via PC running OPUS-NT software, version 6.5. The microscope was equipped with a computer-controlled x,y stage and the sample area within a perspex box was purged in 99,98% pure nitrogen. The Hyperion 3000 microscope was equipped with a mercury cadmium telluride (MCT) based FPA detector of 64x64 pixels where each pixel corresponded to an area of 40x40 µm, so that the visible field corresponded to an area 170x170 µm$^2$ with 15x magnification objective. Each pixel corresponds to a single acquired FT-IR spectrum. For IR measurements, the spectrometer was under vacuum to reduce spectral contributions from water vapour and $CO_2$.

The IR images were continuously acquired with a spectral resolution of 8 cm$^{-1}$ and by co-adding 512 scans of absorbance spectra were recorded in the range from 4000 to 900 cm$^{-1}$. Every cellular sample was measured on $CaF_2$ window in Trasmission mode and a background spectrum/ image of a "blank" area on the same $CaF_2$ window was recorded before each cellular sample measurement in order to account for temporal variations of water vapour and $CO_2$ levels.

Three images were acquired for each $CaF_2$ window, resulting in 30 images with 4096 spectra for the whole experiment. The acquisition time for each FTIR image was approximately 7 minutes.

To study the chemical map after a pre-processing analysis, the spectral data were processed via the rubberband method baseline correction (64 baseline points).

The whole spectral region (4000-900 cm$^{-1}$) was normalized through vector normalization in the spectral range of the Amide I (1595-1800 cm$^{-1}$) band and we have recorded spectra only in the spectral region of *proteins and nucleic acids regions*, from 1800 to 900 cm$^{-1}$. The molecular maps were created from these spectra by plotting different spectral parameters, such as intensities peaks ratio, as a function of x-y pixel position.

*2.5 Data pre-processing for image analysis and statistical analysis*

The resulting FT-IR images were pre-processed by different sofwares (OPUS 6.5, OriginPro 8.0, Mathematica 7.0) and were analysed in proteins and nucleic acids regions (1800-900 cm$^{-1}$). To remove poor quality spectra, the data sets were subjected to a quality test such as *Signal-to-Noise ratio* (S/N) and statistical test such as *Principal Component Analysis* (PCA).

In the (S/N) test, the signal S in each of the 4096 pixels was evaluated as the maximum in the frequency region of the Amide I band (1595-1800 cm$^{-1}$), while the noise N was calculated as the standard deviation (RMS) in the spectral range 1800-1900 cm$^{-1}$.

In order to identify poor quality pixels, those with a Signal-to-Noise ratio S/N smaller than 100 were labeled as *bad pixels* and the corresponding spectra, *bad spectra*.

The threshold set at 100 was chosen in accordance to [9]. The number of bad spectra inside a map is denoted with $N_{bs}$. The maps having more than 1000 bad pixels have been discarded. This acceptance threshold for the number of bad pixels was determined by noting that the dispersion of the spectroscopic parameter $R_1$ (defined later) around the mean value increased significantly for $N_{bs}$ > 1000. In particular, bad spectra tend to populate the low-$R_1$ tail of the distribution, that is thus significantly skewed towards the left for large values of $N_{bs}$. On the contrary, the $R_1$ distribution of good spectra tends to be Gaussian.

We also show how poor spectra can be singled out using PCA. PCA is a multivariate method where a complex data set containing P (*each spectrum of a single map*) different sets of Y variables (*the absorbance at wavenumbers from 1800 to 900 cm$^{-1}$*) is transformed into a smaller set of new indipendent variables or *principal components* (PCs), which maximise the variance of the original data set and are actually unrelated (whereas the original, untransformed variables may have been correlated to some extent).

Therefore, the new reference system identified by the PCs is expressed as a linear combination of the original data set. PCs are computed hierarchically, with the first PC (PC1) accounting for the maximum amount of variance and the others for the subsequent maximum residual variance. The coordinates of the spectra in the new reference system are called *scores* and the coefficients of the linear combination describing each PC, i.e. the weights of the original variables on each PC, are called *loadings*.

After the data pre-processing on the quality of FT-IR maps, a second statistical test known as *Hierarchical Cluster Analysis* (HCA), was introduced to evaluate the chemical similarity between control and sonicated maps of cellular samples; we have extracted the average spectrum of each map for control and sonicated cells, and then the average spectrum was processed with a baseline, smooth and vector normalization at Amide I as described in section 2.4.

To calculate the inter-spectral distance or similarity distance (H) is used the *squared euclidean distance* while the intra-cluster heterogeneity or cluster method (h) is assessed by *Ward's minimum-variance* (OPUS 6.5 software).

In the analysis of FT-IR spectroscopic data combined with cytogenetic-results, we have introduced the *Fisher statistical test* (F-test), which allows to establish if any linear correlation exists between any two variables of the pool.

Finally, the statistical significance in the CBMN assay was evaluated by *Dunnet's test*.

*2.6 Cytokinesis-block micronucleus (CBMN) assay*

The control and sonicated cellular samples were treated with 6 μg/ml *cytochalasin B*. The cells were sampled 24 h after addition of cytochalasin B by centrifugation for 5 min; then, hypotonic treatment consisted of careful resuspension of the cellular samples in 5 ml hypotonic saline (75 mM KCl). Immediately after addition of hypotonic solution, the cells were collected and they fixed in *Carnoy's fixative* (3:1 methanol / acetic acid).

Finally, the cells were transferred onto pre-cleaned slides and were stained with 10 μg/ml of *4'-6'-diamidino-2-phenylindole* (DAPI) in antifade solution (Vector Laboratories).

For each concentration of a test compound 500 binucleated cells (BNC) were evaluated.

Finally, the Nuclear Division Index (NDI), a parameter of cellular mitogen response and cytogenetic effect of US, was evaluated for each sample according to Eastmond and Tucker (1989):

$$NDI = \frac{[M1 + 2(M2) + 3(M3) + 4(M4)]}{N}$$

where $M1$, $M2$, $M3$, $M4$ indicate the number of mono-, bi-, tri- and quadrinucleate cells and $N$ is the total number of counted viable cells. For the scoring of micronuclei the following criteria were adopted from Fenech et al. (2003).

## 3. Results and Discussion

*3.1 Spectra pre-processing*

From the results of the S/N quality test, we have established the number of bad spectra and have evaluated their weight in each data set.

We report, as an example, the distribution of S/N values in two maps, one of the control cells and one of the sonicated cells (SON$_{45\_15}$), that has shown a larger number of bad spectra (Figure 2).

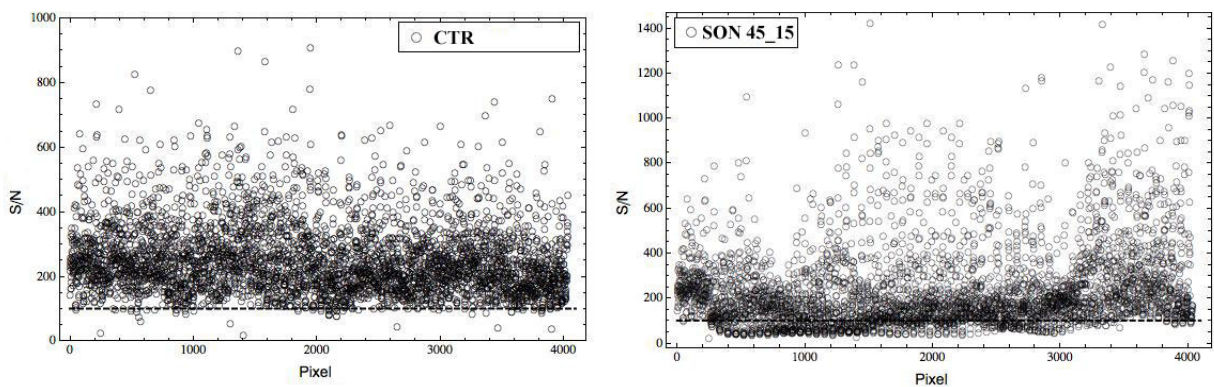

**Figure 2**. *Distribution of the S/N ratio vs. Pixel (0-4096): on the left, control cells (4051 good spectra and 44 bad spectra), and on the right, sonicated cells for 45 minutes at 15 cm (3344 good spectra and 752 bad spectra).*

By Figure 2, we extract a histogram where we also visualize the total number of pixels of the map for control and sonicated ($SON_{45\_15}$) cells corresponding to the values of S/N ratio.

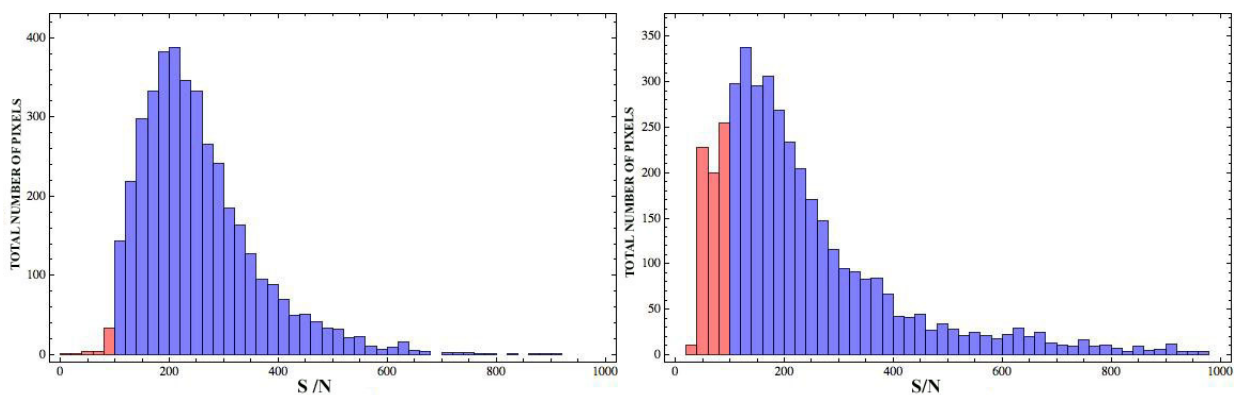

**Figure 3**. *The total number of Pixels corresponding to S/N values: on the left, control cells and on the right, sonicated cells for 45 minutes at 15 cm; the good spectra are shown in blue and the bad spectra are shown in red.*

We have not discarded the FT-IR maps with $N_{bs}<1000$ to monitor the presence of major number of empty spaces due to the US that perturbed the spatial cellular distribution on the slide forming the islands of cellular monolayers.
As shown in Figure 4, we have compared the number of bad spectra vs. time of FT-IR maps between the control and sonicated cells at two distances from the US source.

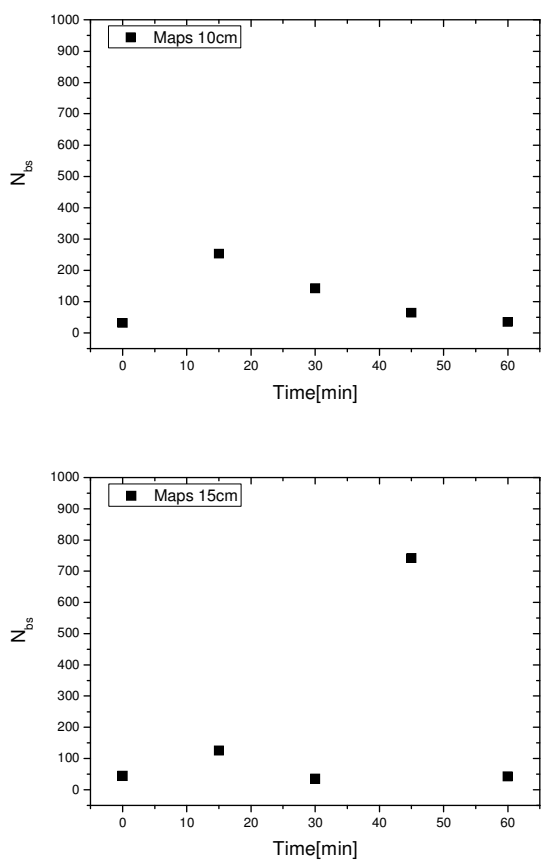

**Figure 4.** *The number of bad spectra ($N_{bs}$) vs. time of US exposure within the maps reported at the 10 and 15 cm from the US source.*

Using the PCA statistical tool, large spectral data were reduced into a small number of indipendent variables known as principal components (PC1, PC2, PC3 etc.); contributions of these components to a given spectra are called *scores*.

The score of the principal components is one of the parameters widely used for classification. In this research, we have employed PCA method to discriminate poor quality spectra from good quality ones within the FT-IR maps.

The PCA eigenvalue plot (data not reported) has shown that the PCs which have the most significant information, were only the first and second principal component (PC1, PC2) for all spectra in the FT-IR maps.

In Figure 5, we have reported an example of PCA results where the information due to the CTR map and $SON_{45\_15}$ map is compressed in the first and second principal component of PCA (see the percentages of the variance in the caption of Figure 5).

The PCA analysis of $SON_{45\_15}$ map shows as the majority of the bad spectra (red points) were distributed towards the negative values of PC2 and were dissimilar from the good spectra (black points); in the CTR map, the bad spectra (red points) were similar between them and were distribuited as the good spectra (black points) towards the positive value of PC1.

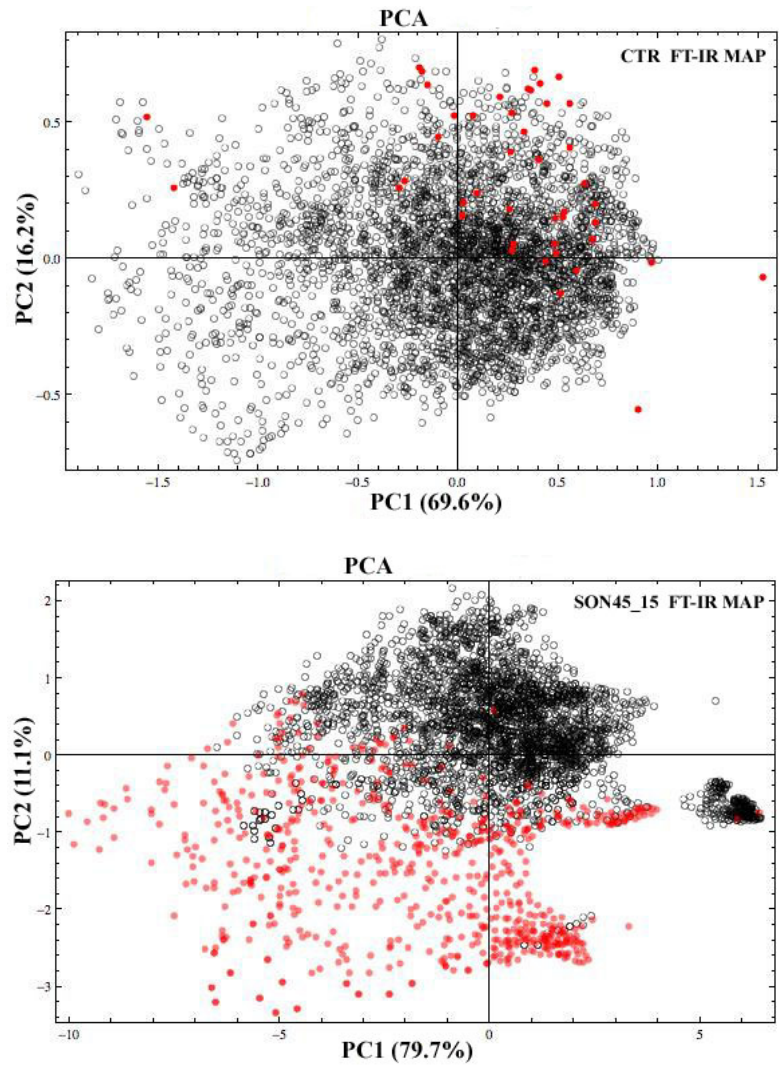

**Figure 5.** *The PCA results (black points=good spectra; red points=bad spectra) obtained for CTR (top) and SON45_15 (bottom) FT-IR map. In the CTR sample, the values of PC1 and PC2 are 69.6% and 16.2%, respectively; in the SON45_15 sample, PCs values are different from CTR sample (PC1=79.7% and PC2=11.1%).*

*3.2 CBMN assay*

Using the cytogenetic test, we have found that the sonicated cells have reported a micronuclei frequency higher than the control cells at both distances.
As described in Figure 6, we have observed a statistically significant value ($p = 0.045$) only for the sonicated cells for 60 minutes at 15 cm.

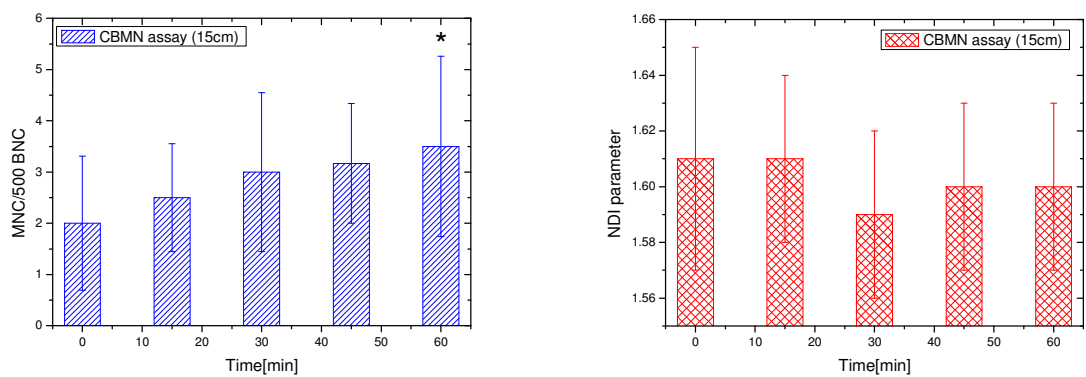

**Figure 6.** *On the left, the histogram describes the value of micronucleated cells (MNC) / 500 binucleated (BNC) cells for control and sonicated cells at 15 cm from the source; significance compared to controls: *p < 0.045. On the right, the histogram shows the NDI parameter vs. time; no significant differences are discovered.*

Finally, no significant differences between the control and sonicated cells for the values of NDI parameter were detected for both distances.

*3.3 FTIR Spectral Imaging*

FTIR spectral imaging enables the determination of the distribution of several molecules of interest; we have studied the chemical map and the average spectrum due to the FT-IR imaging through the analysis of a spectroscopic parameter belonging to the proteins and nucleic acids regions (1800-900 $cm^{-1}$):

- $$R_1 = \frac{A_{1I}}{A_{2I}}$$

where $A_{nI\ (n=1,2)}$ parameters indicate the intensity of Amide I and Amide II peaks respectively and finally, the $R_1$ parameter indicates the ratio due to the intensity of Amide I and Amide II peaks.

The intensities of IR peaks provide quantitative analysis about sample contents, depending on the nature of molecular structure, their bonds, and their environment. The Amide I ($A_1$) and Amide II ($A_2$) bands are centered in the control spectrum at 1653 $cm^{-1}$ and 1543 $cm^{-1}$ respectively. The shape of the Amide I band is influenced by the overall composition in the secondary structure of the samples; the relative contributions fall in the following spectral regions: α-helix between 1645-1662 $cm^{-1}$, β-sheets between 1613-1637 $cm^{-1}$, β-turns between 1662-1682 $cm^{-1}$ and random coil between 1637-1645 $cm^{-1}$.

Before showing the $R_1$ parameter analysis, we have applied the Hierarchical Cluster Analysis method to investigate the relationships among spectra for all samples in the spectral range (1800-900 $cm^{-1}$) at both distances from the US source.

Through this statistical test, the average spectrum extracted from the control and sonicated maps was classified within their respective type cluster according to biochemical similarities.

As described in Figure 7, the maximum heterogeneity level in the proteins and nucleic acids regions was H=1.38 at 15 cm distance; while the maximum heterogeneity within clusters was h=0.14 at both distances.

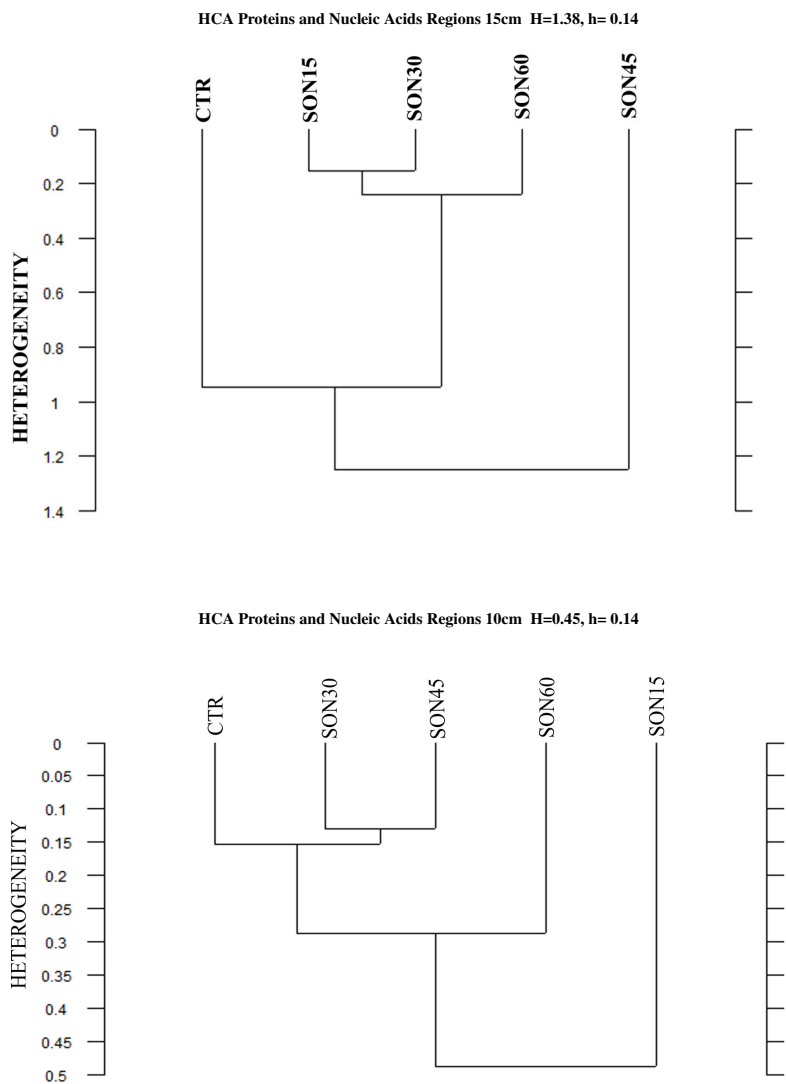

**Figure 7.** *HCA results show the cluster members according to the biochemical similarities for proteins and nucleic acids regions at 10 and 15 cm from the US source.*

The spectral comparison through HCA test is consistent with the results of S/N quality test and PCA analysis obtained from the FT-IR maps; HCA results were also confirmed by the analysis of $R_1$ parameter for all the average spectra.

As shown in Figure 8, we have analysed the Amide I / Amide II (parameter $R_1 = \dfrac{A_{1I}}{A_{2I}}$ ) intensity ratio that represents an indirect measure of DNA content, due to the carbonyl group from the bases and the spectral changes in the range of 1245-960 cm$^{-1}$ that suggests conformational changes and/or rearrangement of existing nucleic acid structures [14].

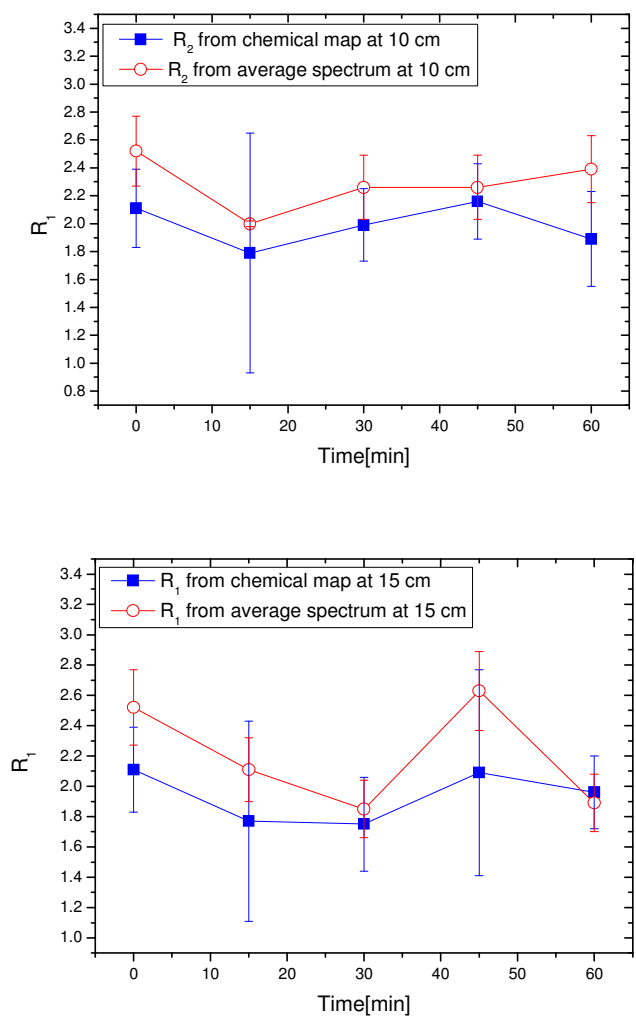

**Figure 8.** *Amide I / Amide II intensity ratio vs. time of US exposure at 10 cm (top) and 15 cm (bottom) from chemical map and average spectrum is reported, showing small differences between the values of chemical map with those of average spectrum.*

The DNA content is detectable by IR spectroscopy mainly when the cells are in the S phase of the cell cycle because the DNA is so tightly packed in the nucleus in the G1 and G2 phases that it appears opaque to IR radiation [15-18].

Assuming that the cells are into the S phase, the small spectral changes in $R_1$ ratio, observed in our experiments and confirmed by HCA for the proteins and nucleic acids regions, could correspond to a better detection of DNA content.

In the following Figure 9, we have reported the comparison between the chemical maps of $R_1$ parameter for CTR and $SON_{45\_15}$ sample, where it is present a considerable difference between the number of bad spectra; moreover, the spectral profiles of CTR and $SON_{45\_15}$ average spectrum were compared.

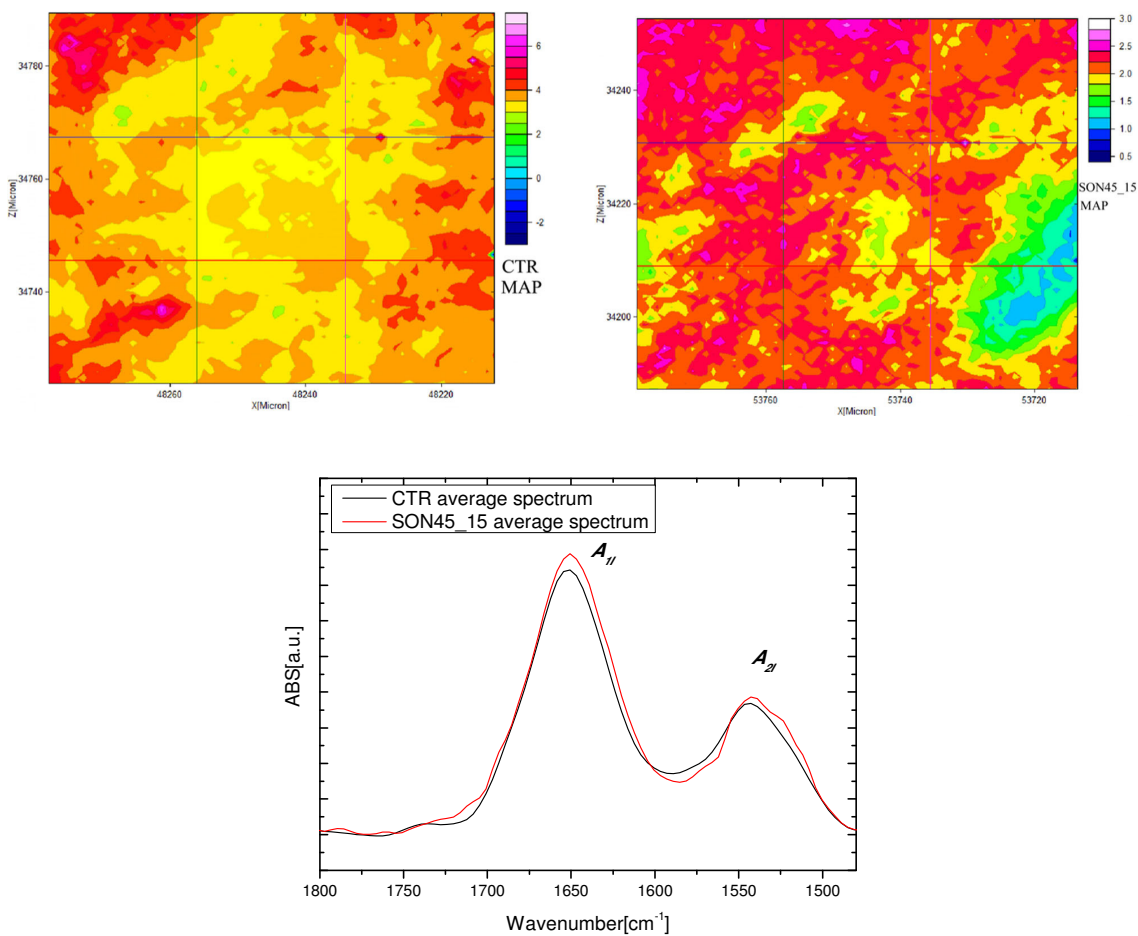

**Figure 9.** *On the top, the chemical maps of $R_1$ parameter for control (left) and sonicated cells for 45 minutes at 15 cm (right); on the bottom, the overlap of average spectrum for CTR and SON45_15 samples, showing the spectral shape of the Amide I and Amide II bands.*

In general, the use of FT-IR imaging or spectroscopy only, without other techniques, does not allow to understand if the nature of the spectral changes in the $R_1$ parameter, due to DNA unpacking effects [19], is linked directly with cell division or the apoptotic process.
Therefore, we have used the micronuclei test to check a correlation between the results of chemical map and the Nuclear Division Index (NDI) parameter.
Figure 10 shows the correlation between the $R_1$ parameter and the NDI at 15 cm distance from the US source; this correlation has a **R**-value of about 0,85 with a probability of about 99% (Fisher test) for the linear significance.
At 10 cm distance from the US source (data not shown), the degree of correlation decreases to a **R**-value of about 0,70 with a probability of about 98% (Fisher test). Thus, the small spectral changes in $R_1$ ratio are coherent with the results of micronuclei test: the Ultrasound does not cause the loss of structural and functional integrity of DNA at both distances of US exposure.

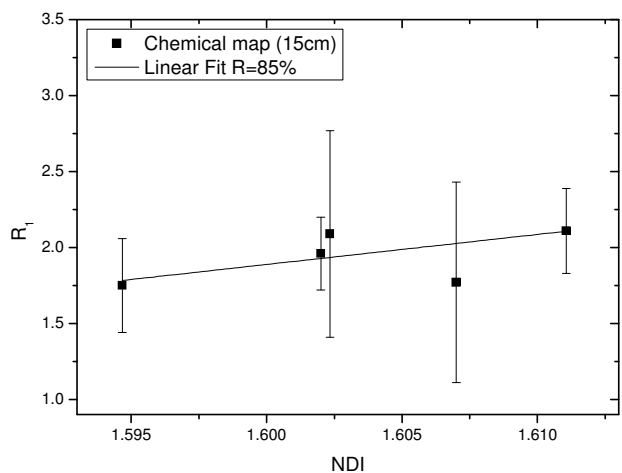

**Figure 10.** *Correlation between the Amide I / Amide II intensity ratio and the NDI parameter with a R-value of about 0,85 and a probability of about 99% (Fisher test).*

## 4. Conclusions

The information obtained in this study can be utilized as experimental basis to detect the ultrasonic effects *in vitro*. Through the results of FT-IR spectral imaging correlated with applied pression and a cytogenetic test, we have found no damage on functionality and structure of DNA ($R_1$ parameter) at 10 and 15 cm from the US source. Therefore, the Ultrasound at 1 MHz frequency does not cause DNA mutations differently from known environmental agents (ultraviolet light, nuclear radiation or certain chemicals). In the pre-processing analysis, the (S/N) ratio was evaluated to determine the number of bad spectra within the maps and the maps with a number of bad spectra over 1000 were discarded. We have also used a second method, PCA, to evaluate the distribution of bad spectra inside the map.

We have used the HCA test to study the chemical similarity between the control and the sonicated samples and we have found correspondence between the results of this test and the variation in $R_1$ parameter, i.e. maps with similar average values of $R_1$ result (1800-900 cm$^{-1}$) to be clustered together. From spectral data in the proteins and nucleic acids regions, we have not found a perfect agreement between the results of the chemical map and the average spectrum. This could be due first to the fact that the average value of $R_1$ is different from the ratio of the average values of $A_{1I}$ and $A_{2I}$. Moreover, the $R_1$-value obtained from the map is actually the mode of the distribution of $R_1$-values, and this concides with the mean only if the distribution is gaussian; this distribution is not gaussian due to the presence of the bad pixels.

Other experiment are in progress to verify whether the ultrasonic source can produce a transient and/or permanent phenomenon of sonoporation on plasma membrane useful for transfection process and to study the size of pores and the membrane recovery time due to different kinetics for small and large molecules *in situ*.

## Acknowledgments

The authors are grateful to Dr. K. Wehbe for their supporting the FTIR-spectral imaging at Beamline B22 FTIR-microspectroscopy of the Diamond Light Source (Oxfordshire). We want also to thank Dr. S. Gaudenzi, Dr. A. Bedini, Dr. C. Giliberti and Dr. R. Palomba for discussions and help in the experiments. Finally, we are grateful to Dr. M. Lattanzi for his assistance with Mathematica software and Dr. S. Belardinelli for his assistance in the laboratory experiments.